# Proton-proton and electron-positron collider in a 100 km ring at Fermilab

C.M. Bhat[1] P.C. Bhat[1], W. Chou[1], E. Gianfelice-Wendt[1], J. Lykken[1], G.L. Sabbi[2], T. Sen[1], R. Talman[3]
Submitted to Snowmass 2013: Frontier Capabilities
June 7, 2013

**Abstract:** The discovery of a Higgs-like boson with mass near 126 GeV, at the LHC, has reignited interest in future energy frontier colliders. We propose here a proton-proton (pp) collider in a 100 km ring, with center of mass (CM) energy of ~100 TeV which would have substantial discovery potential for new heavy particles and new physics beyond the Standard Model. In the case that LHC experiments have already found exotic resonances or heavy "partner" particles, this collider could fill out the "tower" of resonances (thus e.g. confirming an extra dimension) or the full suite of partner particles (e.g. for supersymmetry). The high luminosity of the new collider would enable unique precision studies of the Higgs boson (including Higgs self coupling and rare Higgs decays), and its higher energy would allow more complete measurements of vector boson scattering to help elucidate electroweak symmetry breaking. We also discuss an e+e- collider in the same 100 km ring with CM energies from 90 to 350 GeV. This collider would enable precision electroweak measurements up to the ttbar threshold, and serve as a Higgs factory.

Introduction

The monumental discovery of the long-sought Higgs-like boson [H(126)] at the Large Hadron Collider (LHC) – the last missing piece of the Standard Model (SM) or possibly the harbinger of new physics beyond the SM – has brought excitement to the world high energy physics (HEP) community. Even though the data indicate, so far, the new boson's properties to be consistent with a SM Higgs boson, its measured mass of 126 GeV is also tantalizingly consistent with, for example, a light SM-Higgs-like boson from new physics beyond the SM (BSM), such as supersymmetry (SUSY). It is, therefore, vital to study this new boson in great detail. In addition to upgrading the LHC to increase the integrated luminosity, the HEP community has begun to explore options for a Higgs factory lepton collider [1] to probe the nature of the H(126) particle and for precision measurements of the electroweak sector, complementing the physics at the LHC.

It is possible that the Higgs-like boson is hinting at new physics that might be just around the corner. When the LHC begins its second run in 2015, *pp* collisions at $\sqrt{s} \sim 14$ TeV may reveal new particles and phenomena from BSM physics. We may find signatures of new physics – evidence for SUSY, extra spatial dimensions, new generations of heavy fermions, heavy gauge bosons (W', Z'), or quark compositeness. Since no new physics at $\sqrt{s} \sim 8$ TeV has been detected yet, it is quite likely that we may detect only some of the low-lying states or particles of new physics that are at the limit of reach for LHC energies – e.g. low-lying gauginos but not the squarks; candidates for stop (top squark) but not other super-partners; one or two resonances from a Randall-Sundrum sequence of resonances but others may stay out of reach. Therefore, extending the energy frontier beyond the LHC would be vital to elucidate the nature of electroweak symmetry breaking and whatever new physics is found, and to answer the

[1]: Fermilab, Batavia, IL 60510, [2]: LBNL, Berkeley, CA 94720, [3]: Cornell University, Ithaca, NY 14853

outstanding fundamental questions in particle physics. Historically, significant increases in the energy of hadron machines have often led to important discoveries. A hadron collider at significantly higher collision energies than the LHC, e.g., at $\sqrt{s} \sim$ 100 TeV, would be a facility with substantial potential to discover and explore the new landscape of physics beyond the SM.

In order to know whether H(126) has non-SM behavior, it will be necessary to measure its couplings to fermions and other bosons with very high precision – probably better than 1-2%. In the upcoming LHC runs at $\sqrt{s} \sim$ 14 TeV and with ~300 fb$^{-1}$ data sets, one can measure these couplings to ~5-10% accuracy. With the upgraded high-luminosity LHC (HL-LHC), planned for after 2020, the LHC experiments, with an order of magnitude larger data sets, could measure the couplings to ~2-5%, except for Htt and Higgs self-couplings [1]. Lepton colliders are better suited to make such precision measurements in a cleaner collision environment, while also offering potential for discoveries of rare and/or invisible decays of the Higgs boson. For detailed studies of the H(126) boson and precision electroweak measurements from a CM energy at around the Z boson mass (~90 GeV) to the ttbar threshold (~350 GeV), a circular e+e- collider appears to be an ideal, cost-effective candidate. At $\sqrt{s} \sim$ 250 GeV, the cross section for ZH production is ~200 fb, which would yield 2x10$^4$ ZH events with 100 fb$^{-1}$ of integrated luminosity.

We propose that the US community undertake a design study of a Very Large Hadron Collider (VLHC) to be sited at Fermilab, to create *pp* collisions at 100 TeV and an initial luminosity of ~2x10$^{34}$ cm$^{-2}$s$^{-1}$. With the available dipole magnet technology in the appropriate timeframe, this could require a 100 km circumference tunnel. If plans for an e+e- collider as a Higgs Factory in other parts of the world do not materialize, the 100 km tunnel can also host an e+e- collider to span the CM energies of 90 – 350 GeV. We discuss preliminary design aspects and required R&D for such pp and e+e- colliders in a 100 km ring.

Proton-Proton Collider
*Key characteristics: The main technological components are the superconducting magnets with required dipole fields in the range of 15-20 T. The main operational challenges are likely to be in providing adequate machine protection to cope with the GJs of beam energy and protecting the superconducting components from the high doses of synchrotron radiation.*

A two stage design of the VLHC was considered in 2001 with a low field and a high field stage, both contained in a 233km ring [2]. The low field ring used 2T transmission line magnets to reach a beam energy of 20 TeV. The high field ring used ~10T magnets to reach a beam energy of 87.5 TeV. An e+e- collider, called the VLLC, placed in the same tunnel was also considered during that design phase [3]. Advances in technology over the past decade now make it possible to design colliders in a smaller ring, making them more efficient and possibly cheaper. Below we briefly outline the technology and accelerator design choices and challenges related to this collider.

**Technology** : The superconducting magnets are the crucial determinants of cost and energy reach of this collider. A design based on 15 T arc dipoles makes Nb$_3$Sn the conductor of choice for this application. Nb$_3$Sn magnet technology has made considerable progress in recent years, with the DOE-HEP programs at BNL, FNAL and LBNL playing leadership roles, and more recently through the US-LARP collaboration. In particular, as a result of the US LARP R&D program, Nb$_3$Sn has been selected for upgrading the LHC

Interaction Region quadrupoles for the HL-LHC project [4]. During the next ~10 years, fabrication of full scale prototypes, followed by series production, installation and testing in the LHC will fully validate all aspects of this technology for use in a challenging accelerator environment.

The choice of a 15 T nominal field should be compared with a conductor limited (short sample) field of 17-18 T in optimized $Nb_3Sn$ dipole designs operating at 1.9K. This operating level is generally consistent with those selected for other high-energy accelerators. However, detailed studies taking into account tunnel size, arc filling factors, magnet design and production/operating margins will be required to achieve an optimal balance of performance, cost and risk. Based on previous design studies and the performance of realistic models, the optimal operating field for this application is expected to be in the range of 14 to 15 T. Optimized designs and fully engineered prototypes for this application could be developed over a period of about 10 years through a focused R&D effort similar to the LARP program.

The use of high temperature superconductors such as YBCO and Bi-2212 is being actively explored to surpass the intrinsic limits of $Nb_3Sn$. While these materials can in principle support dipole fields well above 20 T, their application in large accelerators presents significant challenges. These options can be considered for possible later upgrades, rather than as a baseline choice for the collider.

**Example of design parameters**: This design uses similar principles as adopted for the VLHC design in 2001. The rf frequency has been chosen to be an integer multiple of the 53 MHz rf frequency in the Main Injector and Tevatron, to enable easier beam transfers. The bunch intensity is kept low to enable a small emittance. Upgrades to the injectors may allow higher bunch intensities at the same emittance. The number of interactions per bunch crossing is fairly modest, considering that the HL-LHC is planning for an average of ~200 interactions per bunch crossing.

|  | Units | Value |
|---|---|---|
| Circumference | km | 100 |
| Top energy | TeV | 50 |
| Initial Luminosity | $cm^{-2} s^{-1}$ | $2 \times 10^{34}$ |
| Number of IPs |  | 2 |
| Bunch intensity |  | $1.2 \times 10^{10}$ |
| Beam current | mA | 99 |
| Peak dipole field | T | 15.1 |
| Initial normalized emittance (x,y) | μm | 1.0 |
| Initial rms bunch length | cm | 1.9 |
| Initial beam-beam parameters | - | 0.0022, 0.0007 |
| Rf voltage | MV | 80 |
| Harmonic number |  | 215693 |
| Number of bunches |  | 17255 |
| Bunch spacing | ns | 18.8 |
| Average number of interactions per crossing | - | 49 |
| Synchrotron radiation damping time | h | 1.1 |
| Synchrotron radiation power per beam | MW | 0.43 |
| Stored energy per beam | GJ | 1.7 |

**Beam Dynamics** : The challenges associated with this collider are mostly known and much has been learnt from the Tevatron and continuing operation of the LHC. For example, they have reiterated the critical importance of machine protection and beam collimation. In addition to conventional multi-stage collimators, it may be possible to use novel concepts to improve the collimation efficiency. These include bent crystal channeling, hollow electron lenses and rotating collimators.  The amount of beam energy in this collider will be at the level of a few GJ, an order of magnitude higher than in the LHC at design parameters.  Synchrotron radiation will also significantly affect beam dynamics and machine design and operation. Damping times will be of the order of an hour and the synchrotron radiation power will be of the order of a MW. Coping with this radiation level and the consequent beam induced pressure rise will require  R&D in vacuum chamber design and the development of photon stops to protect the superconducting magnets and other equipment.  Accelerator physics experiments planned for the near future are likely to have a significant impact on the design and operation of this collider. We list some of these experiments here.

- A head-on beam-beam compensation experiment at RHIC with electron lenses. If successful, this would allow increasing the transverse brightness of beams and increasing the luminosity.
- A long-range beam-beam compensation with current carrying wires planned at the LHC. A successful test would allow smaller crossing angles, therefore increasing luminosity, and reducing the aperture of interaction region (IR) quadrupoles and also the bunch spacing. Smaller magnet apertures would reduce the cost and shorter bunch spacings would reduce the pile up.
- Use of a crab cavity in each IR with a crossing angle.  If their operation in a future stage of the LHC is successful, it would demonstrate that luminosity loss due to a crossing angle can be recovered.
- A small ring called the Integrable Optics Test Accelerator (IOTA) at the Advanced Superconducting Test Accelerator (ASTA) facility at Fermilab will test the idea of a resonance free lattice. If successful, this will allow reducing the magnet apertures in the arcs and hence significantly reduce cost and in addition make beam operation more stable.
- Application of new cooling techniques such as optical stochastic cooling, coherent electron cooling  or  improvements in microwave stochastic cooling could lower emittances and lead to higher luminosities.

**Injectors** : We envision using Project X as a high intensity low emittance 8 GeV proton pre-accelerator to the existing Main Injector which can accelerate beams to 150 GeV. Another injector will be needed to accelerate beam to 4-5 TeV for injection into the collider. One possibility is to reuse the Tevatron ring (circumference = 6.3 km) with ~15 T magnets, an alternative would be to build a Fermilab site-filler synchrotron  with circumference 16km and lower field magnets.

Electron-Positron Collider
*Key characteristics: The main technological challenges are in supplying the amount of RF power needed, comparable to that for LEP2. The main accelerator design challenges are in combating beamstrahlung and designing a chromaticity correction system which maintains adequate dynamic aperture.*

**Technology**: The rf voltage requirements are similar to those in LEP so one can envisage using similar cavities but with higher power klystrons. Improving the efficiency of conversion of wall plug power to beam power in the rf systems will require R&D with the benefit of limiting the total power consumption for a desired luminosity. The vacuum system needs to cope with a high radiation load and photons in the MeV range. Some R&D may be needed in order to cope with the activation caused by these energetic photons. Dipole magnets in the arcs of the collider require very low fields, in the range of 0.001 – 0.05T in a 100 km ring. Their design can be similar to the ones for LEP but ensuring proper field quality in these magnets will require some R&D. The superconducting quadrupoles in the IR must have large apertures but with relatively modest gradients.

**Example of design parameters**: The following table shows some parameters for an e+e- collider. The design assumes that the rf power available to the beams is kept constant at 100MW. We have made some conservative assumptions in the design, e.g in the filling factors and in the rf acceptance. A lower rf acceptance would enable a lower rf voltage, higher beam current and higher luminosity. This may be possible because a simple calculation suggests that beamstrahlung effects in this ring may not be important below 166 GeV, but it needs confirmation with detailed calculations. Note that the rf voltage even in this conservative design is only about 10% higher than that needed for LEP2. We have assumed one detector, but since the e+e- detector technology is well understood and relatively inexpensive, another interaction point and detector could be added to double the number of Higgs events. The additional chromaticity burden from the second IR will have to be compensated while maintaining sufficient dynamic aperture.

| Parameter | Units | Value |
|---|---|---|
| Energy | GeV | 120 |
| Circumference | km | 100 |
| Beta functions at IP ($\beta_x^*, \beta_y^*$) | cm | 20, 0.2 |
| Number of IPs |  | 1 (or 2) |
| Rf frequency | MHz | 650 |
| Damping partition numbers (Jx, Jy, Jz) |  | 1, 1, 2 |
| Particles/bunch |  | $7.9 \times 10^{11}$ |
| Emittance ($\varepsilon_x, \varepsilon_y$) | [nm] | 16, 0.08 |
| Beam-beam tune shifts |  | 0.095, 0.135 |
| Number of bunches |  | 34 |
| Beam current | mA | 12.9 |
| Energy loss /turn | [GeV] | 1.51 |
| Rf voltage | GV | 3.9 |
| Rf acceptance |  | 0.03 |
| Total Rf power, both beams | MW | 100 |
| Synchrotron rad. power, both beams | MW | 39 |
| Rms bunch length | mm | 3.2 |
| Synchrotron tune |  | 0.223 |
| Bremsstrahlung lifetime | min | 101 |
| Hourglass factor |  | 0.81 |
| Luminosity | $\times 10^{34}$ cm$^{-2}$ s$^{-1}$ | 1.8 |

**Beam Dynamics** : The design of the e+e- collider benefits from the considerable experience gained in operating such colliders over the last few decades. However there are challenges dealing with the radiation emitted by the high energy beams, namely beamstrahlung and synchrotron radiation. In a circular collider, the high energy tail of beamstrahlung can cause particles to be lost from the energy aperture and limit the beam lifetime [5]. Mitigating beamstrahlung requires either i) a sufficiently large energy aperture, typically around 3% or ii) sufficiently rapid top-up injection or iii) increasing the horizontal beam size at the interaction point (IP) to lower the magnetic field of the beam. However this also lowers the luminosity. Better ideas to cope with beamstrahlung without decreasing the luminosity or increasing the cost and complexity should be investigated. One way of increasing luminosity is by lowering beta* at the IPs - the major resulting challenge is to design a chromaticity correction system for the IRs with sufficient energy acceptance and dynamic aperture. Other challenges include mitigating the background in the detectors, controlling the level of synchrotron radiation in the final focus quadrupoles and other machine detector interface issues. Ultra low emittance lattice designs of the type envisaged for future light sources should be investigated and their performance compared regarding chromaticity and dynamic aperture. If applicable, they could lower the emittance by two orders of magnitude. Other beam dynamics issues that need to be studied include the impact of coherent and incoherent beam-beam interactions, synchro-betatron resonances with large synchrotron tunes, and coherent instabilities such as the transverse mode coupling, especially at lower energies in the top-up injector.

**Upgrade path** : We envisage that the collider will first be operated at a beam energy of 120 GeV as a Higgs factory. Later the energy could be increased to 175 GeV with the luminosity decreasing with the cube of the energy if the power stays constant.

**Injectors for the collider** : We envision the top-up injector to exist in the same tunnel as the collider. It would ramp rapidly and continuously from 12 GeV to 120 GeV; suitable fast ramping magnets are in preliminary stages of design [6]. Several scenarios exist for an injector chain with the present Fermilab complex to accelerate the beam to 12 GeV. For the higher energy of 175 GeV, we may need to increase the lower energy of the top up injector and also make appropriate changes in the upstream injectors.